\documentclass[twocolumn]{aastex62}

\DeclareMathAlphabet{\mathsc}{OT1}{cmr}{m}{sc}
\def\testbx{bx}%
\DeclareRobustCommand{\ion}[2]{%
\relax\ifmmode
\ifx\testbx\f@series
{\mathbf{#1\,\mathsc{#2}}}\else
{\mathrm{#1\,\mathsc{#2}}}\fi
\else\textup{#1\,{\mdseries\textsc{#2}}}%
\fi}
\newcommand{\Hei} {\ion{He}{i}}

\newcommand{\msun}{\mbox{M$_{\odot}$}}
\newcommand{\msol}{\mbox{M$_{\odot}$}}

\newcommand{\kms}{\mbox{$\rm{km}\,s^{-1}$}}

\newcommand{\ergs}{erg s$^{-1}$}

\newcommand{\CaII}{Ca~{\sc ii}}

\newcommand{\Nifs}{$^{56}$Ni}
\newcommand{\mej}{$M_\mathrm{ej}$}
\newcommand{\ek}{$E_\mathrm{k}$}
\newcommand{\vph}{$v_\mathrm{ph}$}

\newcommand{\lp}{$L_\mathrm{p}$}
\newcommand{\trise}{$t_{-1/2}$}
\newcommand{\tdecay}{$t_{+1/2}$}

\newcommand{\Emw}{$E\left(B-V\right)_\mathrm{MW}$}

\newcommand{\texp}{$t_\mathrm{exp}$}

\graphicspath{{./}{figures/}}

\received{\today}
\submitjournal{ApJ}

\shorttitle{2018cow}
\shortauthors{Prentice et al.}
\begin{document}

\title{The Cow: discovery of a luminous, hot and rapidly evolving transient}

\correspondingauthor{Simon Prentice}
\email{sipren.astro@gmail.com}

\author[0000-0003-0486-6242]{S. J. Prentice}
\affil{Astrophysics Research Centre, School of Mathematics and Physics, Queen's University Belfast, BT7 1NN, UK\\}
\author[0000-0002-9770-3508]{K. Maguire}
\affil{Astrophysics Research Centre, School of Mathematics and Physics, Queen's University Belfast, BT7 1NN, UK\\}
\author{S. J. Smartt}
\affil{Astrophysics Research Centre, School of Mathematics and Physics, Queen's University Belfast, BT7 1NN, UK\\}
\author{M. R. Magee}
\affil{Astrophysics Research Centre, School of Mathematics and Physics, Queen's University Belfast, BT7 1NN, UK\\}
\author{P. Schady}
\affil{Max-Planck-Institut f{\"u}r Extraterrestrische Physik, Giessenbachstra\ss e, 85748, Garching, Germany\\}
\author{S. Sim}
\affil{Astrophysics Research Centre, School of Mathematics and Physics, Queen's University Belfast, BT7 1NN, UK\\}
\author{T.-W. Chen}
\affil{Max-Planck-Institut f{\"u}r Extraterrestrische Physik, Giessenbachstra\ss e, 85748, Garching, Germany\\}
\author{P. Clark}
\affil{Astrophysics Research Centre, School of Mathematics and Physics, Queen's University Belfast, BT7 1NN, UK\\}
\author{C. Colin}
\affil{Universite de Pierre et Marie Curie (Paris IV), 4 Place Jussieu, 75252, Paris Cedex 5, France}
\affil{Astrophysics Research Centre, School of Mathematics and Physics, Queen's University Belfast, BT7 1NN, UK\\}
\author{M. Fulton}
\affil{Astrophysics Research Centre, School of Mathematics and Physics, Queen's University Belfast, BT7 1NN, UK\\}
\author{O. McBrien}
\affil{Astrophysics Research Centre, School of Mathematics and Physics, Queen's University Belfast, BT7 1NN, UK\\}
\author{D. O'Neill}
\affil{Astrophysics Research Centre, School of Mathematics and Physics, Queen's University Belfast, BT7 1NN, UK\\}
\author{K. W. Smith}
\affil{Astrophysics Research Centre, School of Mathematics and Physics, Queen's University Belfast, BT7 1NN, UK\\}

\author{C. Ashall}
\affil{Department of Physics, Florida State University
77 Chefitan Way, Tallahasee 32304, USA}
\author{K. C. Chambers}
\affil{Institute for Astronomy, University of Hawai'i, 2680 Woodlawn Drive, Honolulu, HI 96822, USA}
\author{L. Denneau}
\affil{Institute for Astronomy, University of Hawai'i, 2680 Woodlawn Drive, Honolulu, HI 96822, USA}
\author{H. A. Flewelling}
\affil{Institute for Astronomy, University of Hawai'i, 2680 Woodlawn Drive, Honolulu, HI 96822, USA}
\author{A. Heinze}
\affil{Institute for Astronomy, University of Hawai'i, 2680 Woodlawn Drive, Honolulu, HI 96822, USA}

\author{T. W.-S. Holoien}
\affil{The Observatories of the Carnegie Institution for Science, 813 Santa Barbara St., Pasadena, CA 91101, USA}

\author{M.~E.~Huber}
\affiliation{Institute for Astronomy, University of Hawai'i, 2680 Woodlawn Drive, Honolulu, HI 96822, USA}

\author{C. S. Kochanek}
\affiliation{Department of Astronomy, The Ohio State University, 140 W. 18th Ave., Columbus, OH 43210, USA}
\affiliation{Center for Cosmology and AstroParticle Physics, The Ohio State University, 191 W. Woodruff Ave., Columbus, OH 43210, USA}

\author{P. A. Mazzali}
\affil{Astrophysics Research Institute, Liverpool John Moores University, IC2, Liverpool Science Park, 146 Brownlow Hill,  Liverpool, L3 5RF, UK}
\affil{Max-Planck-Institut f{\"u}r Astrophysik, Karl-Schwarzschild-Str. 1, D-85748 Garching, Germany\\}

\author[0000-0003-1072-2712]{J.~L.~Prieto}
\affil{N\'ucleo de Astronom\'ia de la Facultad de Ingenier\'ia, Universidad Diego Portales, Av. Ej\'ercito 441, Santiago, Chile}
\affil{Millennium Institute of Astrophysics, Santiago, Chile}

\author{A. Rest}
\affiliation{Space Telescope Science Institute, 3700 San Martin Drive, Baltimore, MD 21218, USA}
\affiliation{Department of Physics and Astronomy, Johns Hopkins University, Baltimore, MD 21218, USA}

\author[0000-0003-4631-1149]{B.~J.~Shappee}
\affiliation{Institute for Astronomy, University of Hawai'i, 2680 Woodlawn Drive, Honolulu, HI 96822, USA}

\author{B. Stalder}
\affil{LSST, 950 N Cherry Ave, Tucson, AZ 95719}

\author{K.~Z.~Stanek}
\affiliation{Department of Astronomy, The Ohio State University, 140 W. 18th Ave., Columbus, OH 43210, USA}
 \author[0000-0002-5571-1833]{M. D. Stritzinger}
 \affil{Department of Physics and Astronomy, Aarhus University, Ny Munkegade 120, DK-8000 Aarhus C, Denmark}

\author{T. A. Thompson}
\affiliation{Department of Astronomy, The Ohio State University, 140 W. 18th Ave., Columbus, OH 43210, USA}
\affiliation{Center for Cosmology and AstroParticle Physics, The Ohio State University, 191 W. Woodruff Ave., Columbus, OH 43210, USA}

\author{J. L. Tonry}
\affil{Institute for Astronomy, University of Hawai'i, 2680 Woodlawn Drive, Honolulu, HI 96822, USA}

% \author[0000-0002-0786-7307]{August Muench}
% \affiliation{American Astronomical Society \\
% 2000 Florida Ave., NW, Suite 300 \\
% Washington, DC 20009-1231, USA}
% \collaboration{(AAS Journals Data Scientists collaboration)}



\begin{abstract}

We present the ATLAS discovery and initial analysis of the first 18 days of the unusual transient event, ATLAS18qqn/AT2018cow. It is characterized by a high peak luminosity ($\sim$1.7 $\times$ 10$^{44}$ erg s$^{-1}$), rapidly evolving light curves 
($>$5\,mag rise to peak in $\sim$3.5 days), 
and hot blackbody spectra, peaking at $\sim$27000 K that are relatively featureless and unchanging over the first two weeks. The bolometric light curve cannot be powered by radioactive decay under realistic assumptions.  The detection of high-energy emission may suggest a central engine as the powering source. Using a magnetar model, we estimated an ejected mass of $0.1-0.4$ \msol, which lies between that of low-energy core-collapse events and the kilonova, AT2017gfo.
The spectra cooled rapidly from 27000 to 15000 K in just over 2 weeks but remained smooth and featureless. 
Broad and shallow emission lines appear after about 20 days, and we tentatively identify them as  
\Hei\ although they would be redshifted from their rest wavelengths. We rule out that there are any features in the spectra due to intermediate mass elements up to and including the Fe-group. 
The presence of r-process elements cannot be ruled out. If these lines are due to He, then we suggest a low-mass star with residual He as a potential progenitor. Alternatively, models of magnetars formed in neutron-star mergers give plausible matches to the data.\\
\end{abstract}

\section{Introduction} \label{sec:intro}

The advent of wide-field transient surveys that scan the visible sky every few nights has led to the discovery of new classes of transients, such as superluminous supernovae \citep[SLSNe, e.g.,][]{2011Natur.474..487Q}, Type Iax SNe \citep[e.g.][]{2003PASP..115..453L},  and Ca-rich transients \citep[e.g.][]{2010Natur.465..322P}. In particular, high-cadence surveys have uncovered a new parameter space of SN-like events that rise and fall much faster than standard SNe \citep[e.g.][]{Drout2014,2016ApJ...819...35A,2016ApJ...819....5T,2018arXiv180304869P,2018NatAs...2..307R}.  The first confirmed kilonova (AT2017gfo) from a neutron star merger detected in 
gravitational waves (GW170817) is the fastest declining astrophysical transient \citep{2017ApJ...848L..12A} that also approaches SN-like luminosities.

These newly discovered rapidly evolving transients have a wide range of peak absolute magnitudes ($-$15 $>$ M $>$ $-$22\,mag), rise times ($\sim$1--10 days), and spectral properties that make them difficult to explain through a single progenitor scenario but most are incompatible with a radioactively-powered explosion. Proposed scenarios include SN shock breakout in a surrounding wind \citep[e.g.][]{2010ApJ...724.1396O}, cooling low-mass envelopes after a SN shock breakout \citep{2010ApJ...725..904N,2014MNRAS.438..318K}, a magnetar-powered binary neutron star merger \citep{2013ApJ...771...86G,2013ApJ...776L..40Y,2014MNRAS.439.3916M}, and an optical flare from a tidal disruption event \citep{2009MNRAS.400.2070S}.  

In this Letter, we report the discovery of the unusual, luminous, and fast-evolving transient, ATLAS18qqn/AT2018cow (nicknamed `The Cow') discovered by the ATLAS survey \citep[][]{2018PASP..130f4505T}. We present initial observations from ultra-violet (UV) to near-infrared (NIR) wavelengths out to $\sim$18-24 days post discovery. AT2018cow was also detected in the X-ray, radio, and sub-millimeter \citep[e.g.,][]{2018MNRAS.tmpL.149R,2018ATel11749....1D} but these observations are not the focus of this paper.

\section{Observations and Data analysis} \label{sec:obs}
ATLAS is a twin 0.5-m telescope system installed on the Hawai'ian islands of Haleakala and Mauna Loa \citep[][]{2018PASP..130f4505T}. Each unit has a 28.9 square degree field of view that is robotically surveying the sky in cyan (\textit{c}) and orange (\textit{o}) filters that are broadly equivalent to Pan-STARRS/Sloan Digital Sky Survey (SDSS) \textit{g+r} and \textit{r+i} filters, respectively. ATLAS typically covers the whole sky visible from Hawaii every two nights. We discovered a new transient, ATLAS18qqn, 
in a 30-second exposure with start time 2018-06-16 10:35:38 UT Modified Julian Date (MJD) 58285.44141 at an AB magnitude of $o=14.74\pm0.10$. It was assigned the International Astronomical Union\footnote{https://wis-tns.weizmann.ac.il/} name AT2018cow and announced as an unusual transient by \cite{2018ATel11727....1S}. 

\begin{figure}
	\centering
    \includegraphics[scale=0.35]{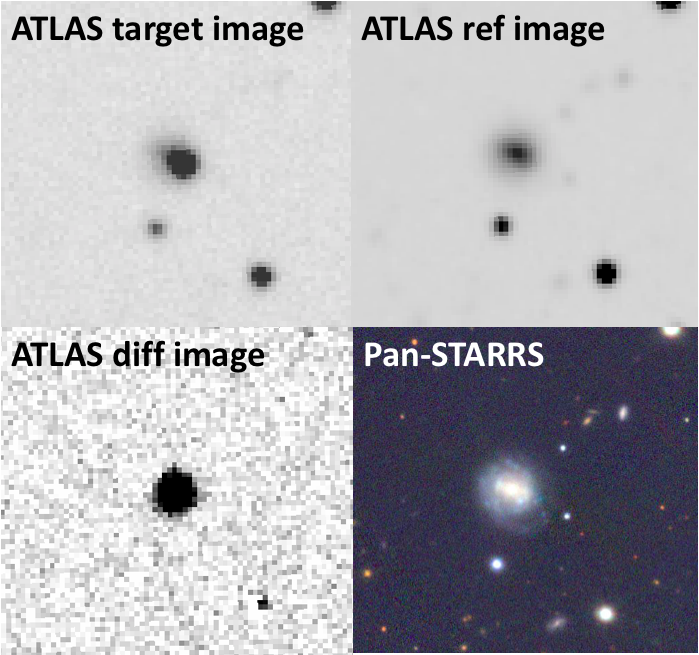}
    \caption{Images showing the location of AT2018cow: a post-discovery image (top left), a pre-discovery reference image (top right), a subtracted difference image (bottom left) and a Pan-STARRS multi-color image (bottom right). }
    \label{fig:atlas_disc}
\end{figure}

AT2018cow is offset by 1.7 kpc from the core of the galaxy CGCG 137-068 (Figure \ref{fig:atlas_disc}). A SDSS DR6 spectrum \citep[][]{2008ApJS..175..297A,2013AJ....146...32S} shows the galaxy to be star-forming with nebular emission lines at a redshift of $z=0.014145$. A cosmology with $H_0=73$ km s$^{-1}$ Mpc $^{-1}$, $\Omega_m=0.27$, $\Omega_\mathrm{\Lambda}=0.73$, gives a 
distance of 66$\pm$5 Mpc.   We corrected for Milky Way (MW) extinction of \Emw\ $=0.08$\,mag \citep{2011ApJ...737..103S} using a \cite{1989ApJ...345..245C} R$_V$=3.1 extinction law. 
We assume that the host galaxy extinction is negligible.

ATLAS did not detect the source 3.95 days\footnote{Times denoted in ``days'' are observer frame, those in ``d'' in rest-frame.} prior to the first detection to a depth of $>$20.2\,mag (3$\sigma$ limit in \textit{o} band). The All Sky Automated Survey for SuperNovae \citep[ASAS-SN,][]{2014ApJ...788...48S} did not detect the source to a depth of $>$18.9\,mag (3$\sigma$ limit in \textit{g} band) just 1.3 days before the ATLAS detection and robustly detected the source
3 days later (see Table\,\ref{tab:phot}). The explosion epoch was estimated from modelling of the bolometric light curve (see Section \ref{bolo}) to be MJD 58284.3, with the ASAS-SN non-detection 0.2 days before.

\subsection{Light curves}
We began monitoring AT2018cow starting 1.7 days after discovery \citep{2018ATel11729....1C,2018ATel11734....1C} in \textit{g'r'i'z'JHK} with GROND \citep{2008PASP..120..405G} on the 2.2-m MPG telescope and then in $ugriz$ with IO:O on the Liverpool Telescope \citep[LT;][]{2004SPIE.5489..679S} beginning 4.6 days after discovery. The optical and NIR data were calibrated using SDSS and 2MASS stars \citep{2008ApJ...685..376K}.  Observations with the UV Optical Telescope \citep[UVOT;][]{2005SSRv..120...95R} on board the Neil Gehrels Swift Observatory \citep{2004ApJ...611.1005G} were also obtained and were calibrated using standard procedures \citep{2008MNRAS.383..627P}.

\begin{figure*}
	\centering
	\includegraphics[scale=0.4]{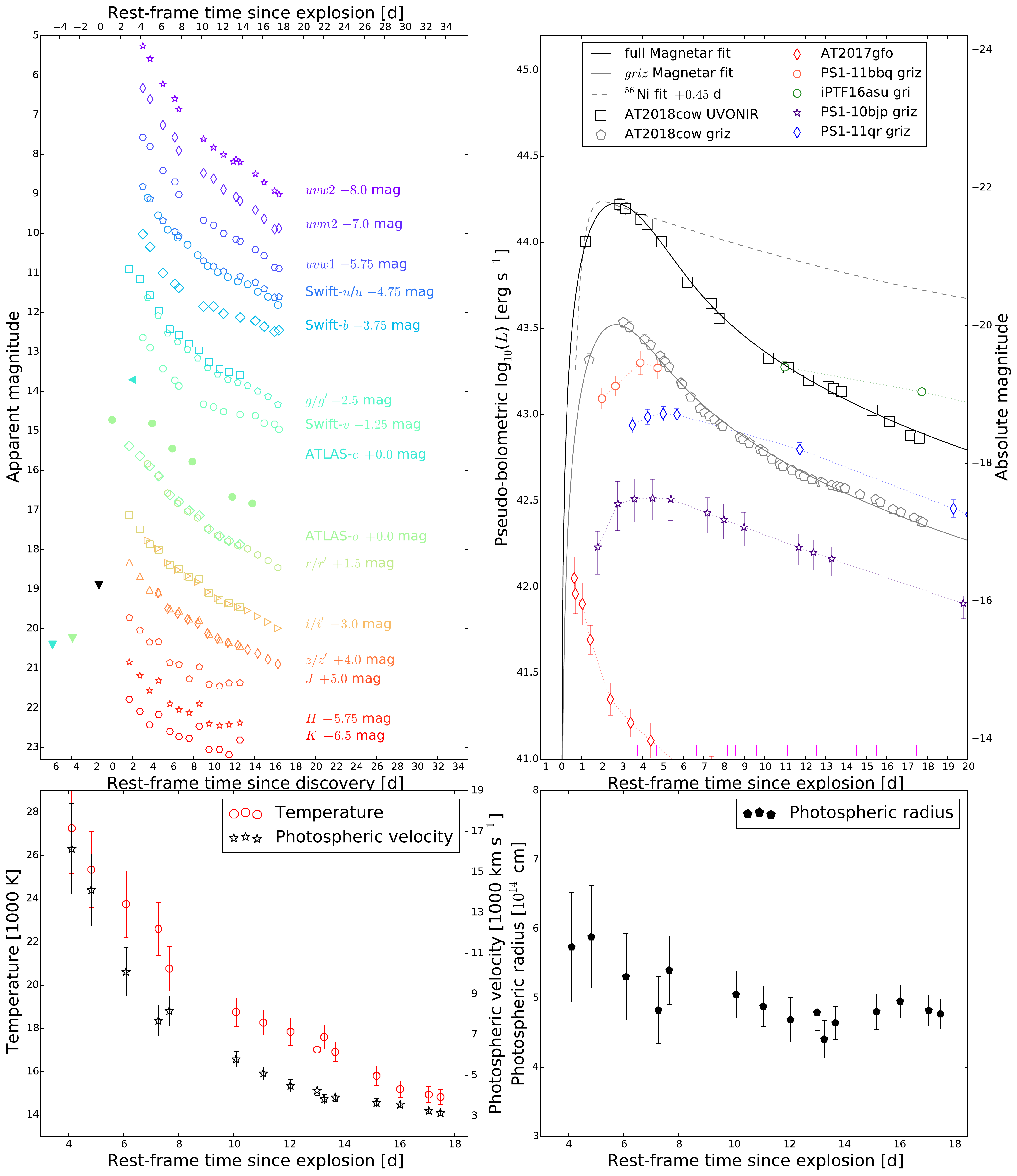}

     \caption{{\bf Upper left panel:} The ATLAS, LT, GROND, and SWIFT light curves of AT2018cow. The observations are in the rest frame, with \texp\ estimated from the light curve models. The ASAS-SN non-detection (black down-arrow) is shown, along with the last ATLAS non-detections. {\bf Upper right panel:} The UVONIR (black) and $griz$ (grey) bolometric light curves of AT2018cow, the dotted line is the time of the ASAS-SN non-detection. Other luminous transients with short rise times are also shown: iPTF16asu \citep{Whitesides2017}, PS1-11bbq, PS1-bjp, PS1-11qr \citep{Drout2014}, and the kilonova AT2017gfo \citep{Smartt2017}. No K-corrections have been applied to the photometry. Magnetar model fits to AT2018cow are shown as a black/grey solid lines, while the best-fitting \Nifs-powered model is shown as a grey dashed line. Magenta lines along the bottom indicate the dates of spectral observations. {\bf Lower left panel:} The effective temperature and velocity evolution of AT2018cow. {\bf Lower right panel:} The photospheric radius as function of time.}
    \label{fig:18cowLC}
\end{figure*}

Fig.~\ref{fig:18cowLC} shows the multi-color light curves of AT2018cow. Maximum light occurred on MJD 58286.9 (from the light curve models, see Section \ref{bolo}). An ATLAS data point obtained $+$0.6 days from maximum has $m_c =$13.6\,mag ($-$20.5\,mag, uncorrected for MW extinction).  After peak, the light curves decayed at a rate of $0.05-0.2$\,mag per day, with the bluer bands typically decaying faster than the redder bands.  A flattening is seen in the light curves  at $\sim$2 weeks after discovery. This flattening can be seen most clearly in the right-hand panel of Fig.~\ref{fig:18cowLC}, where there is an excess in the \textit{griz} bolometric light curve luminosity compared to the smooth magnetar fit (see Section \ref{bolo}). The \textit{g}-band decline between peak and 15 days post maximum is $\Delta m_{15}$($g$) $\approx 3$\,mag.

%%%%%%%%%%%%%%%%%%%%%%%SPECTROSCOPY%%%%%%%%%%%%%%%%%%%%%%%%%%%
\subsection{Spectroscopy} \label{sec:spec}
\begin{figure*}
	\centering
    \includegraphics[scale=0.8]{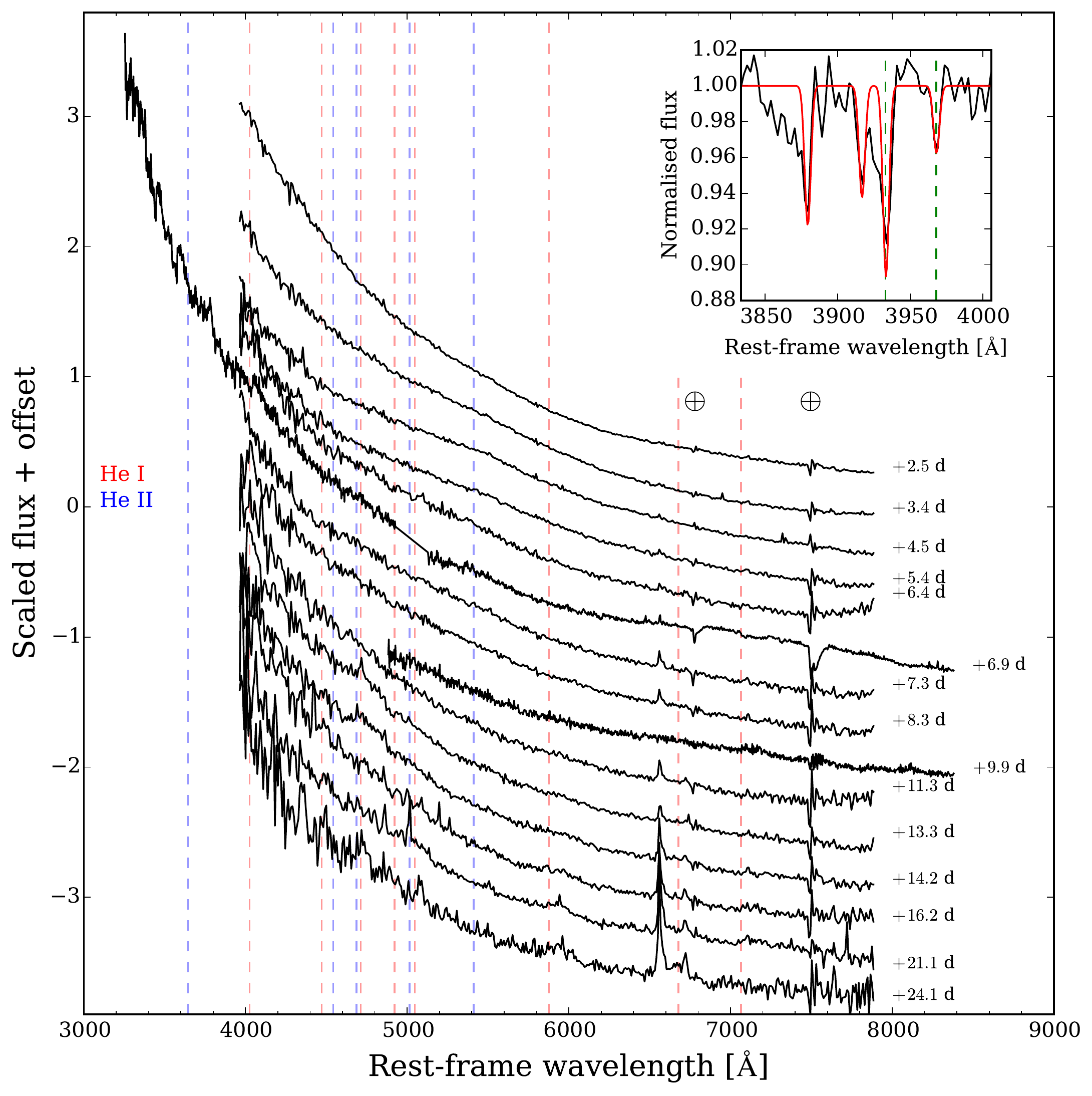}
    \caption{Spectra of AT2018cow (LT, ACAM, SNIFS) to $+24.1$ d, all epochs are rest-frame time since detection with the first spectrum at approximately maximum light. The inset shows the host \CaII\ H\&K lines at $z=0.0139$ (green dashed line). }
    \label{fig:18cowspectra}
\end{figure*}

Optical spectroscopy was obtained at the LT with the low-resolution ($R\sim350$) SPRAT spectrograph starting $2.6$ days after discovery (Fig.~\ref{fig:18cowspectra}). The spectra were reduced using the standard LT pipeline \citep{2012AN....333..101B} and a custom {\sc python} script.
Subsequent spectra ($R=1000$ in the blue) were obtained at the University of Hawaii 2.2-m telescope with SNIFS \citep{2002SPIE.4836...61A} and at the 4.2-m William Herschel Telescope (WHT) with ACAM \citep{2008SPIE.7014E..6XB} using a 0.5\arcsec\ slit to obtain $R\sim700$. The spectra were flux calibrated to coeval photometry.

The SNIFS spectrum shows narrow absorption features from the MW and the host galaxy, CGCG 127-68. We measured two redshifted components of the Ca II H\&K lines to give a consistent redshift of $z=0.0139$.  

Analysis of the 2D spectral frames revealed that the narrow emission feature consistent with the rest wavelength of H$\alpha$ is extended  leading us to conclude the emission feature in these early spectra results from the host galaxy and not AT2018cow. 

The early-time spectra are blue and quite featureless, as first suggested by \cite{2018ATel11732....1P}.  Little evolution is seen in the spectra up to $\sim$2 weeks after explosion,  apart from a decrease in temperature. It was initially suggested that AT2018cow was spectroscopically similar to a broad-line Type Ic SN after subtraction of a power-law component \citep{2018ATel11740....1X,2018ATel11753....1I}. 
We find that the power-law subtracted spectra of AT2018cow do not evolve similarly to the spectra of GRB-SN 1998bw \citep[e.g.][]{Galama1998,2001ApJ...555..900P}. 
\cite{2018arXiv180800969P} also present an extensive follow-up data set of AT2018cow, finding 
similar conclusions.

At 21.1 d after detection ($t=22.2$ d), the spectra of AT2018cow started to show broad features in the wavelength range 5900 -- 6100 \AA. To aid in line identification, we calculated a series of model spectra using TARDIS, a one-dimensional Monte Carlo radiative transfer code \citep{TARDIS,TARDIS-zenodo}.   The features could be emission of \ion{He}{ii}~$\lambda$4686, \ion{He}{i}~$\lambda$5015 or \ion{He}{ii}~$\lambda$5005, \ion{He}{i}~$\lambda$5876, and \ion{He}{i}~$\lambda$6678, respectively (see Figure \ref{fig:18cowspectra}) but the potential emission features appear redshifted with respect to the rest position by $\sim 3000$ \kms.  The identification of these features with \ion{He}{i} and \ion{He}{ii} but offset to the red was also suggested by \cite{2018ATel11836....1B}.

%%%%%%%%%%%%%%%%
% Temperature and photospheric velocity
%%%%%%%%%%%%%%%%
\section{Temperature and photospheric velocity evolution}
The initial temperature of AT2018cow was estimated by modelling the spectral energy distribution as a blackbody to be $27000\pm{2000}$ K at $t=4.1$ d (Figure~\ref{fig:18cowLC}). The temperature then shows a progressive decline over the next two weeks to $\sim 15000$ K.

Assuming homologous expansion and that AT2018cow was spherical, the photospheric velocity, \vph, and photospheric radius, $R_\mathrm{ph}$, were also estimated (Figure~\ref{fig:18cowLC}). The velocity at $t=4.1$ d is \vph\ $\sim 16000\pm{2000}$ \kms, declining to $\sim 3000$ \kms\ in two weeks. Over the same period, the photospheric radius stays relatively constant at $\sim 5\times10^{14}$ cm.

%%%%%%%%%%%%%%%%%%%%%%%%
% Bolometric light curve
%%%%%%%%%%%%%%%%%%%%%%%%

\section{Bolometric light curve analysis}
\label{bolo}
Figure~\ref{fig:18cowLC} shows the pseudo-bolometric (henceforth ``bolometric") light curve of AT2018cow. It was constructed using the UV to NIR photometry (UVONIR, $1850 - 23000$ \AA) and the method described in \cite{Prentice2016}. Spline fits to the light curves were used to interpolate the fluxes on {\it SWIFT} observation dates. 

AT2018cow reached a peak UVONIR luminosity, \lp\ $\approx 1.7\times 10^{44}$ \ergs\ ($M = -21.8$\,mag). Measurements of the characteristic light-curve timescales using a spline fit to the data gives a rise time from \lp/2 to \lp\ of \trise\ $<1.7$ d and an equivalent decay time after peak of \tdecay\ $=2.5\pm{0.5}$ d. Constraints from the photometric non-detections give a limit on the rise time of $<3.3$ d to increase $>$5\,mag.

%%%%%%%%%%%%%%%%%%%%%%%%%%MODEL FITS%%%%%%%%%%%%%%%%%%%%%%
\subsection{Model fits to the bolometric light curves}
The best-fitting \Nifs-powered light curve model \citep{Arnett1982,2008MNRAS.383.1485V} has a $^{56}$Ni mass of $\sim$ 3 \msol\ and 0.05 -- 0.3 \msun\ of ejecta (for realistic ejecta velocities), which is unphysical. This model fits the peak luminosity and the rise, but not the decay (Figure~\ref{fig:18cowLC}), and no model fits all three.

We also investigated powering of AT2018cow by the highly-magnetized rapidly rotating neutron star (magnetar) models of \cite{Kasen2010} as formulated in \cite{Inserra2013}. For our model, we assumed spherical symmetry and 100$\%$ efficiency in thermalizing the spin-down energy.  The best fitting model (Figure~\ref{fig:18cowLC}) has a spin period, P  $\approx 11$ ms, a magnetic field strength, B $\approx 2.0\times10^{15}$ G, an explosion time, $t_{\rm exp}$ $\approx 1.1$ d before the ATLAS discovery, and a rise time to maximum light of $t_{rise}$ $\approx 2.5$ d. The model fit to the $griz$ light curve gives similar timescales but with $P\approx 25$ ms and $B\approx 3.5\times10^{15}$ G.

Using $t_{rise}$, and assuming an opacity of $0.1 - 0.2$ cm$^{2}$ g$^{-1}$ and a kinetic energy in the range, $10^{51}$ $<$ \ek $<$ $10^{52}$ erg, we estimated a ejecta mass, \mej\ $= 0.1 - 0.4$ \msol\ for the magnetar model. This lies in between the \mej\ of the kilonova, AT2017gfo \citep[\mej\ = 0.04$\pm$0.01 \msol;][]{Smartt2017,2017Sci...358.1570D} and low-mass stripped-envelope core-collapse events such as SN 1994I \citep[\mej\ $\sim 1$ \msol;][]{1994Natur.371..227N}.

Late-time accretion onto a central compact object is predicted to roughly follow a t$^{-5/3}$ decay law \citep[e.g.][]{1989ApJ...346..847C}, which is similar to the t$^{-2}$ used in the magnetar model. Therefore, a fallback accretion scenario \citep{2013ApJ...772...30D} for AT2018cow predicts a similar \mej.

%%%%%%%%%%%%%%%%%%%%%%%%
\section{Discussion and conclusions} \label{sec:disc}
%%%%%%%%%%%%%%%%%%%%%%%%
The combination of its high peak luminosity ($\sim$1.7 $\times$ 10$^{44}$ erg s$^{-1}$), fast rise time ($>$5\,mag in 3.3 d), high peak blackbody temperature ($\sim$27000 K), low ejecta mass (0.1 -- 0.4 \msol), and relatively featureless and non-evolving spectra make AT2018cow very unusual. Some analogues at higher redshift may exist \citep{Drout2014,2018arXiv180304869P}, but discovery of events like AT2018cow are unprecedented in the local Universe. 

A key result of our analysis is that a magnetar or accretion model requires a low ejecta mass of $\sim0.1-0.4$ \msol, which is between that of a low-mass core-collapse event and the kilonova, AT2017gfo.  

From our spectral analysis, we tentatively identify emission lines of \Hei. The peaks of the emission features are not quite aligned with the rest frame \Hei\ wavelengths. They are 
redshifted, suggestive of a large bulk velocity for the He-rich material. The presence of He is difficult to reconcile with either magnetar or accretion models since such a progenitor should have previously lost all its He.

Models such as shock-breakout or recombination in an extended envelope that have been put forward for other fast and luminous events \cite[e.g,][]{Drout2014}.
The shock breakout of SN 1993J 
was nearly two magnitudes fainter than AT2018cow and required a radius of $4\times 10^{13}$ cm, already close to the limit for observed red supergiants \cite[from calculations of][]{1994ApJ...429..300W}. Therefore, an unfeasibly large and extended envelope would be required to power the light curve of AT2018cow via shock breakout.
No signs of narrow line emission consistent with interaction with H/He-rich material is seen for AT2018cow, making a shock breaking out of circumstellar material such as in \cite{2010ApJ...724.1396O} unlikely. 

A number of models have been put forward for the special case of the formation of a magnetar in a binary neutron star merger \citep{2013ApJ...771...86G,2013ApJ...776L..40Y,2014MNRAS.439.3916M}. These magnetar models predict transients that are more luminous and slower evolving than kilonovae (that would occur in addition to the kilonova event). In particular, the \mej\ $=0.1$ \msol\  model with a magnetic field of 10$^{15}$ G of \cite{2014MNRAS.439.3916M} predicts a UV/optical transient with a similar peak luminosity, decline rate, and effective temperature to that of AT2018cow. Although published models are not a perfect match, better fits may be possible by tuning model parameters. This model also predicts non-thermal X-ray emission on a similar timescale to the UV/optical emission.

Multiple X-ray \citep[e.g.,][]{2018MNRAS.tmpL.149R} and radio/sub-millimeter \citep[e.g.,][]{2018ATel11749....1D} detections have been made of AT2018cow.  Further modelling and observations across the full electromagnetic spectrum will hopefully allow the origin of this unusual transient to be determined.

\section{Acknowledgements}
SP, KM and MM are funded by the EU H2020 ERC grant no.~758638; SJS, KS by 
STFC Grants  ST/P000312/1 and ST/N002520/1; 
MDS by the Villum FONDEN (grant no.~13261); TWC by Alexander von Humboldt Foundation. Funding for GROND was generously granted from the Leibniz-Prize to Prof. Hasinger (DFG grant HA 1850/28-1). CSK and KZS are supported by NSF grants AST-1515876 and AST-1515927. ATLAS acknowledges NASA grant NN12AR55G. 
The Liverpool Telescope is operated by Liverpool John Moores University with financial support from the UK STFC. The WHT is operated on La Palma by the Isaac Newton Group of Telescopes. We thank the WHT staff for their prompt scheduling and execution of this target.
We thank the Las Cumbres Observatory and its staff for its continuing support of the ASAS-SN project.  ASAS-SN is supported by the Gordon and Betty Moore Foundation (GBMF5490) and NSF grant AST-1515927. Development of ASAS-SN has been supported by NSF grant AST-0908816, the Mt. Cuba Astronomical Foundation, the Ohio State University, the Chinese Academy of Sciences South America Center for Astronomy (CASSACA), the Villum Foundation, and George Skestos.

\vspace{5mm}
\facilities{LT (IO:O, SPRAT), ATLAS}

\begin{longtable*}{lccccc}

	\caption{UVONIR photometry of AT2018cow.}\\
	
    \hline
    MJD & Phase$^a$ & $m$ & $\delta m$ & Filter & Telescope \\
    \hline
58279.50 & $-5.6$  & $>20.4$ &   &  $c$ & ATLAS\\
58281.48 & $-3.9$  & $>20.2$ &  & $o$ & ATLAS \\
58284.13 & $-1.3$ & $>18.9$ &  & $g$ & ASASSN\\
58285.44 & 0.0 & 14.7 & 0.1 & $o$ & ATLAS \\
58287.15 & 1.6 & 13.40 & 0.05 & $g'$ & GROND \\
58287.15 & 1.6 & 13.8 & 0.1 & $r'$ & GROND \\
58287.15 & 1.6 & 14.1 & 0.1 & $i'$ & GROND \\
58287.15 & 1.6 & 14.32 & 0.05 & $z'$ & GROND \\
58287.15 & 1.6 & 14.71 & 0.07 & $J$ & GROND \\
58287.15 & 1.6 & 15.10 & 0.08 & $H$ & GROND \\
58287.15 & 1.6 & 15.3 & 0.1 & $K$ & GROND \\
58287.44 & 1.9 & 13.60 & 0.01 & $c$ & ATLAS \\
58288.20 & 2.7 & 13.65 & 0.05 & $g'$ & GROND \\
58288.20 & 2.7 & 14.1 & 0.1 & $r'$ & GROND \\
58288.20 & 2.7 & 14.4 & 0.1 & $i'$ & GROND \\
58288.20 & 2.7 & 14.67 & 0.05 & $z'$ & GROND \\
58288.20 & 2.7 & 15.04 & 0.07 & $J$ & GROND \\
58288.20 & 2.7 & 15.43 & 0.07 & $H$ & GROND \\
58288.20 & 2.7 & 15.6 & 0.1 & $K$ & GROND \\
58288.50 & 3.0 & 13.25 & 0.03 & $uvw2$ & UVOT \\
58288.50 & 3.0 & 13.32 & 0.04 & $uvm2$ & UVOT \\
58288.50 & 3.0 & 13.31 & 0.04 & $uvw1$ & UVOT \\
58288.50 & 3.0 & 13.56 & 0.06 & Swift-$u$ & UVOT \\
58288.50 & 3.0 & 13.77 & 0.06 & Swift-$b$ & UVOT \\
58288.50 & 3.0 & 13.88 & 0.07 & Swift-$v$ & UVOT \\
58288.98 & 3.4 & 13.84 & 0.08 & $u$ & LT \\
58288.98 & 3.4 & 14.12 & 0.06 & $g$ & LT \\
58288.98 & 3.4 & 14.32 & 0.03 & $r$ & LT \\
58288.98 & 3.4 & 14.76 & 0.03 & $i$ & LT \\
58289.17 & 3.6 & 14.06 & 0.05 & $g'$ & GROND \\
58289.17 & 3.6 & 14.3 & 0.1 & $r'$ & GROND \\
58289.17 & 3.6 & 14.8 & 0.1 & $i'$ & GROND \\
58289.17 & 3.6 & 15.01 & 0.05 & $z'$ & GROND \\
58289.17 & 3.6 & 15.34 & 0.07 & $J$ & GROND \\
58289.17 & 3.6 & 15.81 & 0.08 & $H$ & GROND \\
58289.17 & 3.6 & 15.9 & 0.1 & $K$ & GROND \\
58289.22 & 3.7 & 13.57 & 0.06 & $uvw2$ & UVOT \\
58289.22 & 3.7 & 13.60 & 0.07 & $uvm2$ & UVOT \\
58289.22 & 3.7 & 13.55 & 0.07 & $uvw1$ & UVOT \\
58289.22 & 3.7 & 13.87 & 0.07 & Swift-$u$ & UVOT \\
58289.22 & 3.7 & 14.08 & 0.07 & Swift-$b$ & UVOT \\
58289.22 & 3.7 & 14.14 & 0.07 & Swift-$v$ & UVOT \\
58289.42 & 3.9 & 14.70 & 0.06 & $o$ & ATLAS \\
58290.02 & 4.5 & 14.29 & 0.03 & $u$ & LT \\
58290.02 & 4.5 & 14.58 & 0.06 & $g$ & LT \\
58290.02 & 4.5 & 14.63 & 0.03 & $r$ & LT \\
58290.02 & 4.5 & 14.97 & 0.03 & $i$ & LT \\
58290.02 & 4.5 & 15.09 & 0.05 & $z$ & LT \\
58290.08 & 4.5 & 14.45 & 0.05 & $g'$ & GROND \\
58290.08 & 4.5 & 14.6 & 0.1 & $r'$ & GROND \\
58290.08 & 4.5 & 14.9 & 0.1 & $i'$ & GROND \\
58290.08 & 4.5 & 15.08 & 0.05 & $z'$ & GROND \\
58290.08 & 4.5 & 15.33 & 0.08 & $J$ & GROND \\
58290.08 & 4.5 & 15.57 & 0.08 & $H$ & GROND \\
58290.08 & 4.5 & 15.7 & 0.1 & $K$ & GROND \\
58290.50 & 4.9 & 14.22 & 0.09 & $uvw2$ & UVOT \\
58290.50 & 4.9 & 14.2 & 0.1 & $uvm2$ & UVOT \\
58290.50 & 4.9 & 14.16 & 0.07 & $uvw1$ & UVOT \\
58290.50 & 4.9 & 14.42 & 0.07 & Swift-$u$ & UVOT \\
58290.50 & 4.9 & 14.75 & 0.07 & Swift-$b$ & UVOT \\
58290.50 & 4.9 & 14.67 & 0.07 & Swift-$v$ & UVOT \\
58290.97 & 5.4 & 14.65 & 0.03 & $u$ & LT \\
58290.97 & 5.4 & 15.02 & 0.06 & $g$ & LT \\
58290.97 & 5.4 & 15.06 & 0.03 & $r$ & LT \\
58290.97 & 5.4 & 15.33 & 0.03 & $i$ & LT \\
58290.97 & 5.4 & 15.49 & 0.03 & $z$ & LT \\
58291.20 & 5.6 & 14.92 & 0.05 & $g'$ & GROND \\
58291.20 & 5.6 & 15.1 & 0.1 & $r'$ & GROND \\
58291.20 & 5.6 & 15.3 & 0.1 & $i'$ & GROND \\
58291.20 & 5.6 & 15.49 & 0.05 & $z'$ & GROND \\
58291.20 & 5.6 & 15.86 & 0.08 & $J$ & GROND \\
58291.20 & 5.6 & 16.15 & 0.08 & $H$ & GROND \\
58291.20 & 5.6 & 16.1 & 0.1 & $K$ & GROND \\
58291.43 & 5.9 & 15.34 & 0.03 & $o$ & ATLAS \\
58291.69 & 6.1 & 14.59 & 0.06 & $uvw2$ & UVOT \\
58291.69 & 6.1 & 14.57 & 0.07 & $uvm2$ & UVOT \\
58291.69 & 6.1 & 14.4 & 0.1 & $uvw1$ & UVOT \\
58291.69 & 6.1 & 14.70 & 0.07 & Swift-$u$ & UVOT \\
58291.69 & 6.1 & 15.02 & 0.08 & Swift-$b$ & UVOT \\
58291.69 & 6.1 & 14.96 & 0.08 & Swift-$v$ & UVOT \\
58291.98 & 6.4 & 14.85 & 0.07 & $u$ & LT \\
58291.98 & 6.4 & 15.24 & 0.06 & $g$ & LT \\
58291.98 & 6.4 & 15.32 & 0.03 & $r$ & LT \\
58291.98 & 6.4 & 15.51 & 0.03 & $i$ & LT \\
58291.98 & 6.4 & 15.62 & 0.03 & $z$ & LT \\
58292.09 & 6.5 & 14.8 & 0.1 & $uvw2$ & UVOT \\
58292.09 & 6.5 & 14.90 & 0.07 & $uvm2$ & UVOT \\
58292.09 & 6.5 & 14.76 & 0.08 & $uvw1$ & UVOT \\
58292.09 & 6.5 & 14.81 & 0.09 & Swift-$u$ & UVOT \\
58292.09 & 6.5 & 15.12 & 0.08 & Swift-$b$ & UVOT \\
58292.09 & 6.5 & 15.10 & 0.08 & Swift-$v$ & UVOT \\
58292.10 & 6.5 & 15.07 & 0.05 & $g'$ & GROND \\
58292.10 & 6.5 & 15.2 & 0.1 & $r'$ & GROND \\
58292.10 & 6.5 & 15.4 & 0.1 & $i'$ & GROND \\
58292.10 & 6.5 & 15.54 & 0.05 & $z'$ & GROND \\
58292.10 & 6.5 & 15.97 & 0.08 & $J$ & GROND \\
58292.10 & 6.5 & 16.30 & 0.09 & $H$ & GROND \\
58292.10 & 6.5 & 16.2 & 0.1 & $K$ & GROND \\
58292.96 & 7.4 & 15.0 & 0.1 & $u$ & LT \\
58292.96 & 7.4 & 15.43 & 0.06 & $g$ & LT \\
58292.96 & 7.4 & 15.52 & 0.03 & $r$ & LT \\
58292.96 & 7.4 & 15.66 & 0.03 & $i$ & LT \\
58292.96 & 7.4 & 15.76 & 0.04 & $z$ & LT \\
58293.12 & 7.5 & 15.28 & 0.05 & $g'$ & GROND \\
58293.12 & 7.5 & 15.5 & 0.1 & $r'$ & GROND \\
58293.12 & 7.5 & 15.6 & 0.1 & $i'$ & GROND \\
58293.12 & 7.5 & 15.74 & 0.05 & $z'$ & GROND \\
58293.12 & 7.5 & 16.12 & 0.08 & $J$ & GROND \\
58293.12 & 7.5 & 16.37 & 0.08 & $H$ & GROND \\
58293.12 & 7.5 & 16.3 & 0.1 & $K$ & GROND \\
58293.43 & 7.8 & 15.67 & 0.01 & $o$ & ATLAS \\
58293.97 & 8.4 & 15.3 & 0.07 & $u$ & LT \\
58293.97 & 8.4 & 15.65 & 0.06 & $g$ & LT \\
58293.97 & 8.4 & 15.69 & 0.03 & $r$ & LT \\
58293.97 & 8.4 & 15.82 & 0.03 & $i$ & LT \\
58293.97 & 8.4 & 15.86 & 0.04 & $z$ & LT \\
58294.13 & 8.5 & 15.45 & 0.05 & $g'$ & GROND \\
58294.13 & 8.5 & 15.6 & 0.1 & $r'$ & GROND \\
58294.13 & 8.5 & 15.7 & 0.1 & $i'$ & GROND \\
58294.13 & 8.5 & 15.77 & 0.05 & $z'$ & GROND \\
58294.13 & 8.5 & 16.0 & 0.1 & $J$ & GROND \\
58294.13 & 8.5 & 16.15 & 0.09 & $H$ & GROND \\
58294.13 & 8.5 & 16.0 & 0.1 & $K$ & GROND \\
58294.55 & 8.9 & 15.6 & 0.1 & $uvw2$ & UVOT \\
58294.55 & 8.9 & 15.4 & 0.1 & $uvm2$ & UVOT \\
58294.55 & 8.9 & 15.41 & 0.07 & $uvw1$ & UVOT \\
58294.55 & 8.9 & 15.44 & 0.08 & Swift-$u$ & UVOT \\
58294.55 & 8.9 & 15.59 & 0.08 & Swift-$b$ & UVOT \\
58294.55 & 8.9 & 15.57 & 0.09 & Swift-$v$ & UVOT \\
58294.95 & 9.3 & 15.57 & 0.08 & $u$ & LT \\
58294.95 & 9.3 & 15.90 & 0.06 & $g$ & LT \\
58294.95 & 9.3 & 15.97 & 0.03 & $r$ & LT \\
58294.95 & 9.3 & 16.07 & 0.03 & $i$ & LT \\
58294.95 & 9.3 & 16.12 & 0.04 & $z$ & LT \\
58295.10 & 9.5 & 15.75 & 0.05 & $g'$ & GROND \\
58295.10 & 9.5 & 15.9 & 0.1 & $r'$ & GROND \\
58295.10 & 9.5 & 16.1 & 0.1 & $i'$ & GROND \\
58295.10 & 9.5 & 16.12 & 0.05 & $z'$ & GROND \\
58295.10 & 9.5 & 16.4 & 0.1 & $J$ & GROND \\
58295.10 & 9.5 & 16.66 & 0.09 & $H$ & GROND \\
58295.10 & 9.5 & 16.6 & 0.1 & $K$ & GROND \\
58295.55 & 9.9 & 15.8 & 0.1 & $uvw2$ & UVOT \\
58295.55 & 9.9 & 15.61 & 0.08 & $uvm2$ & UVOT \\
58295.55 & 9.9 & 15.5 & 0.1 & $uvw1$ & UVOT \\
58295.55 & 9.9 & 15.58 & 0.08 & Swift-$u$ & UVOT \\
58295.55 & 9.9 & 15.59 & 0.09 & Swift-$b$ & UVOT \\
58295.55 & 9.9 & 15.6 & 0.1 & Swift-$v$ & UVOT \\
58295.95 & 10.3 & 15.7 & 0.7 & $u$ & LT \\
58295.95 & 10.3 & 16.06 & 0.06 & $g$ & LT \\
58295.95 & 10.3 & 16.14 & 0.03 & $r$ & LT \\
58295.95 & 10.3 & 16.23 & 0.03 & $i$ & LT \\
58295.95 & 10.3 & 16.24 & 0.03 & $z$ & LT \\
58296.15 & 10.5 & 15.92 & 0.05 & $g'$ & GROND \\
58296.15 & 10.5 & 16.1 & 0.1 & $r'$ & GROND \\
58296.15 & 10.5 & 16.2 & 0.1 & $i'$ & GROND \\
58296.15 & 10.5 & 16.27 & 0.05 & $z'$ & GROND \\
58296.15 & 10.5 & 16.5 & 0.1 & $J$ & GROND \\
58296.15 & 10.5 & 16.70 & 0.09 & $H$ & GROND \\
58296.15 & 10.5 & 16.6 & 0.1 & $K$ & GROND \\
58296.55 & 10.9 & 16.0 & 0.1 & $uvw2$ & UVOT \\
58296.55 & 10.9 & 15.88 & 0.08 & $uvm2$ & UVOT \\
58296.55 & 10.9 & 15.75 & 0.08 & $uvw1$ & UVOT \\
58296.55 & 10.9 & 15.7 & 0.1 & Swift-$u$ & UVOT \\
58296.55 & 10.9 & 15.78 & 0.08 & Swift-$b$ & UVOT \\
58296.55 & 10.9 & 15.75 & 0.09 & Swift-$v$ & UVOT \\
58296.98 & 11.3 & 15.85 & 0.08 & $u$ & LT \\
58296.98 & 11.3 & 16.19 & 0.06 & $g$ & LT \\
58296.98 & 11.3 & 16.28 & 0.03 & $r$ & LT \\
58296.98 & 11.3 & 16.36 & 0.04 & $i$ & LT \\
58296.98 & 11.3 & 16.35 & 0.06 & $z$ & LT \\
58297.09 & 11.4 & 16.00 & 0.05 & $g'$ & GROND \\
58297.09 & 11.4 & 16.2 & 0.1 & $r'$ & GROND \\
58297.09 & 11.4 & 16.3 & 0.1 & $i'$ & GROND \\
58297.09 & 11.4 & 16.34 & 0.05 & $z'$ & GROND \\
58297.09 & 11.4 & 16.4 & 0.1 & $J$ & GROND \\
58297.09 & 11.4 & 16.67 & 0.09 & $H$ & GROND \\
58297.09 & 11.4 & 16.7 & 0.1 & $K$ & GROND \\
58297.43 & 11.8 & 16.5 & 0.2 & $o$ & ATLAS \\
58297.53 & 11.9 & 16.1 & 0.1 & $uvw2$ & UVOT \\
58297.79 & 12.1 & 16.1 & 0.1 & $uvw2$ & UVOT \\
58297.79 & 12.1 & 16.0 & 0.1 & $uvm2$ & UVOT \\
58297.79 & 12.1 & 15.8 & 0.1 & $uvw1$ & UVOT \\
58297.97 & 12.3 & 15.96 & 0.03 & $u$ & LT \\
58297.97 & 12.3 & 16.27 & 0.06 & $g$ & LT \\
58297.97 & 12.3 & 16.38 & 0.03 & $r$ & LT \\
58297.97 & 12.3 & 16.44 & 0.04 & $i$ & LT \\
58297.97 & 12.3 & 16.41 & 0.04 & $z$ & LT \\
58298.18 & 12.5 & 16.09 & 0.05 & $g'$ & GROND \\
58298.18 & 12.5 & 16.3 & 0.1 & $r'$ & GROND \\
58298.18 & 12.5 & 16.4 & 0.1 & $i'$ & GROND \\
58298.18 & 12.5 & 16.42 & 0.05 & $z'$ & GROND \\
58298.18 & 12.5 & 16.3 & 0.1 & $J$ & GROND \\
58298.18 & 12.5 & 16.63 & 0.09 & $H$ & GROND \\
58298.18 & 12.5 & 16.3 & 0.1 & $K$ & GROND \\
58298.20 & 12.5 & 16.2 & 0.1 & $uvw2$ & UVOT \\
58298.20 & 12.5 & 16.1 & 0.1 & $uvm2$ & UVOT \\
58298.20 & 12.5 & 15.94 & 0.08 & $uvw1$ & UVOT \\
58298.20 & 12.5 & 15.83 & 0.08 & Swift-$u$ & UVOT \\
58298.20 & 12.5 & 15.87 & 0.08 & Swift-$b$ & UVOT \\
58298.20 & 12.5 & 15.83 & 0.09 & Swift-$v$ & UVOT \\
58298.95 & 13.3 & 16.0 & 0.1 & $u$ & LT \\
58298.95 & 13.3 & 16.34 & 0.06 & $g$ & LT \\
58298.95 & 13.3 & 16.47 & 0.04 & $r$ & LT \\
58298.95 & 13.3 & 16.54 & 0.05 & $i$ & LT \\
58298.95 & 13.3 & 16.52 & 0.04 & $z$ & LT \\
58299.40 & 13.7 & 16.7 & 0.1 & $o$ & ATLAS \\
58299.72 & 14.0 & 16.49 & 0.06 & $uvw2$ & UVOT \\
58299.72 & 14.0 & 16.40 & 0.08 & $uvm2$ & UVOT \\
58299.72 & 14.0 & 16.1 & 0.1 & $uvw1$ & UVOT \\
58299.72 & 14.0 & 15.98 & 0.08 & Swift-$u$ & UVOT \\
58299.72 & 14.0 & 15.96 & 0.08 & Swift-$b$ & UVOT \\
58299.72 & 14.0 & 15.85 & 0.09 & Swift-$v$ & UVOT \\
58299.96 & 14.3 & 16.2 & 0.1 & $u$ & LT \\
58299.96 & 14.3 & 16.49 & 0.07 & $g$ & LT \\
58299.96 & 14.3 & 16.63 & 0.04 & $r$ & LT \\
58299.96 & 14.3 & 16.68 & 0.05 & $i$ & LT \\
58299.96 & 14.3 & 16.62 & 0.05 & $z$ & LT \\
58300.59 & 14.9 & 16.7 & 0.1 & $uvw2$ & UVOT \\
58300.59 & 14.9 & 16.63 & 0.09 & $uvm2$ & UVOT \\
58300.59 & 14.9 & 16.31 & 0.08 & $uvw1$ & UVOT \\
58300.59 & 14.9 & 16.1 & 0.1 & Swift-$u$ & UVOT \\
58300.59 & 14.9 & 16.10 & 0.09 & Swift-$b$ & UVOT \\
58300.59 & 14.9 & 16.0 & 0.1 & Swift-$v$ & UVOT \\
58300.98 & 15.3 & 16.35 & 0.03 & $u$ & LT \\
58300.98 & 15.3 & 16.62 & 0.06 & $g$ & LT \\
58300.98 & 15.3 & 16.77 & 0.03 & $r$ & LT \\
58300.98 & 15.3 & 16.83 & 0.03 & $i$ & LT \\
58300.98 & 15.3 & 16.77 & 0.04 & $z$ & LT \\
58301.64 & 15.9 & 16.9 & 0.1 & $uvw2$ & UVOT \\
58301.64 & 15.9 & 16.89 & 0.09 & $uvm2$ & UVOT \\
58301.64 & 15.9 & 16.60 & 0.08 & $uvw1$ & UVOT \\
58301.64 & 15.9 & 16.36 & 0.08 & Swift-$u$ & UVOT \\
58301.64 & 15.9 & 16.23 & 0.09 & Swift-$b$ & UVOT \\
58301.64 & 15.9 & 16.0 & 0.1 & Swift-$v$ & UVOT \\
58301.97 & 16.3 & 16.56 & 0.07 & $u$ & LT \\
58301.97 & 16.3 & 16.82 & 0.07 & $g$ & LT \\
58301.97 & 16.3 & 16.95 & 0.05 & $r$ & LT \\
58301.97 & 16.3 & 16.99 & 0.06 & $i$ & LT \\
58301.97 & 16.3 & 16.89 & 0.05 & $z$ & LT \\
58302.07 & 16.4 & 17.01 & 0.07 & $uvw2$ & UVOT \\
58302.07 & 16.4 & 16.87 & 0.09 & $uvm2$ & UVOT \\
58302.07 & 16.4 & 16.64 & 0.09 & $uvw1$ & UVOT \\
58302.07 & 16.4 & 16.35 & 0.09 & Swift-$u$ & UVOT \\
58302.07 & 16.4 & 16.19 & 0.09 & Swift-$b$ & UVOT \\
58302.07 & 16.4 & 16.2 & 0.1 & Swift-$v$ & UVOT \\
\hline
\multicolumn{6}{l}{$^a$Rest-frame with respect to first observation}\\
	\label{tab:phot}
\end{longtable*}




\bibliographystyle{aasjournal}
\bibliography{astro.bib}



\end{document}